\documentclass{paradigmplus}
\usepackage{amsmath}
\usepackage{tcolorbox} % For colored boxes
\newlength\mylen
\newcommand\myinput[1]{%
  \settowidth\mylen{\KwIn{}}%
  \setlength\hangindent{\mylen}%
  \hspace*{\mylen}#1}

    \title[A Comprehensive System Architecture for Emergency Response in Smart Cities]{A Comprehensive System Architecture using Field Programmable Gate Arrays Technology, Dijkstra's Algorithm, and Edge Computing for Emergency Response in Smart Cities}

    \author[1]{Mahamat Abdel Aziz Assoul \orcid{0009-0005-1038-5017}}
    \author[2]{Abakar Mahamat Tahir \orcid{0000-0001-7715-6458}}
    \author[3]{Taibi Mahmoud \orcid{0000-0001-7032-8579}}
    \author[4]{Garrik Brel Jagho Mdemaya \orcid{0000-0003-2399-8371}}
    \author[4,\correspondingAuthor]{Milliam Maxime Zekeng Ndadji \orcid{0000-0002-0417-5591}}
    
    \affil[1]{Department of Industrial Engineering and Maintenance, Polytechnic University of Mongo, Mongo, Chad \protect\\ \email{mahamatassoulfils@gmail.com}} 

    \affil[2]{Department of Technical Sciences, University of N'Djamena, N'Djamena, Chad \protect\\ \email{abakarmt@gmail.com}} 

    \affil[3]{Department of Electronic, Badji-Mokhtar University of Annaba, Annaba, Algeria \protect\\ \email{mahmoud.taibi@uniiv-annaba.dz}} 
    
    \affil[4]{Department of Mathematics and Computer Science, University of Dschang, Dschang, Cameroon \protect\\ \email{\{jaghobrel@gmail.com, ndadji.maxime@univ-dschang.org\}}}
    
    
    \graphicspath{{assets/}}
    
    \vol{5}
    \no{2}
    \year{2024}
    \doi{10.55969/paradigmplus.v5n2a1}
    \publicationdates{20 April 2024}{31 July 2024}{2 August 2024}
\begin{document}
    \begin{abstract}
       Efficient emergency response systems are crucial for smart cities. But their implementation is highly challenging, particularly in regions like Chad where infrastructural constraints are prevalent. The urgency for optimized response times and resource allocation in emergency scenarios is magnified in these contexts, yet existing solutions often assume robust infrastructure and uninterrupted connectivity, which is not always available. Most of the time, they are based on system architectures pre-designed for other purposes. This paper addresses these critical challenges by proposing a comprehensive system architecture that integrates Field Programmable Gate Arrays (FPGAs), Dijkstra's algorithm, and Edge Computing. The objective is to enhance emergency response through accelerated route planning and resource allocation, addressing gaps left by traditional cloud-based systems. Methodologically, key characteristics of the desired system are identified, then its components are described and their integration is explained; the system leverages FPGA-based computations and a distributed implementation of Dijkstra's algorithm to compute the shortest paths rapidly, while Edge Computing ensures decentralized and resilient processing. A theoretical analysis highlights promising improvements in response times and resource management. The proposed system architecture not only enhances emergency response efficiency but is also adaptable to infrastructural constraints of Chadian-like environments.
    \end{abstract}

%%%%%%%%%%%%%%%%%%%%%%%%%%%%%%%%%%%%%%%%%%%%%%%%%%%%%%%
\section{Introduction}
\label{Introduction}
In the contemporary urban landscape, the evolution of smart cities promises heightened efficiency and enhanced quality of life \cite{REJEB2023100721}. Central to this vision is the optimization of various urban systems, including emergency response services \cite{djikstra_accident_2019, nigerian2017}. Efficient emergency response systems are critical, particularly within the context of smart cities where population density and infrastructural complexity magnify the risks and impacts of emergencies \cite{elsevier2014}. Rapid and effective response to emergencies can significantly mitigate the adverse effects of disasters, saving lives and reducing economic losses.

However, traditional emergency response systems often struggle to meet the demands of these modern urban environments \cite{Cui2022Anovel}. These systems typically rely on centralized cloud computing, which can introduce latency and dependency on continuous network connectivity \cite{emergency_system2022}. This reliance becomes particularly problematic in regions like Chad, where infrastructure limitations, geographic constraints, and socioeconomic factors present unique hurdles. For instance, frequent network interruptions and limited bandwidth can severely hamper the effectiveness of cloud-based solutions in real-time emergency scenarios. Additionally, existing solutions often neglect the optimization of emergency response times and resource allocation, particularly through advanced computational techniques and hardware acceleration. For example, in the event of a natural disaster, it is very common for emergency services to arrive either very late or not at all. This delay may result from late alerts, suboptimal route planning, or complete failures in the alert system \cite{emergency_system2022}.

The efficacy of emergency response systems is contingent upon factors such as time efficiency, resource allocation, and usability within Chadian-like contexts. Existing works often falter in addressing these imperatives adequately. The delay in response times, inefficient resource allocation, and the inability to function effectively under suboptimal network conditions underscore the need for innovative and contextually relevant solutions. 
Motivated by these gaps, this paper proposes a novel system architecture specifically designed to address the critical needs of emergency response in smart cities, with a particular focus on Chadian-like environments. The proposed architecture integrates Field Programmable Gate Arrays (FPGAs), Dijkstra's algorithm, and Edge Computing to enhance emergency response efficiency. The primary contribution of this work is developing a bespoke system architecture that leverages FPGA technology to accelerate the execution of Dijkstra's algorithm for rapid route planning and resource optimization. This approach ensures that the most efficient routes are identified swiftly, even in complex urban environments. The proposed architecture implements Edge Computing to decentralize computational resources, thereby enhancing responsiveness and resilience to network disruptions. By processing data closer to the source, edge computing reduces latency and dependency on central cloud infrastructure, making the system more robust against connectivity issues. In addition, the proposed architecture takes into account the fact that network connectivity may not be good and therefore makes proper use of cache memories to promote and accelerate offline calculations. All the components of the architecture are described and the key algorithms are presented. By addressing these aspects, the proposed architecture aims to set a new benchmark for emergency response systems in smart cities, particularly those with infrastructural challenges similar to Chad. The integration of advanced computational techniques and decentralized processing represents a significant advancement over traditional systems, promising improved efficiency, reliability, and adaptability in emergency situations.

The present document is organized as follows: Section \ref{sec:litterature_review} reviews current work addressing emergency service response in smart cities and discusses the performance of FPGA boards in computing. Section \ref{sec:contribution} presents this paper's contribution by describing the proposed system architecture, detailing its core components, and discussing the FPGA-based implementation of Dijkstra's algorithm. A discussion highlights the strengths and weaknesses of the proposed architecture. Finally, Section \ref{sec:conclusion} concludes the paper.
%%%%%%%%%%%%%%%%%%%%%%%%%%%%%%%%%%%%%%%%%%%%%%%%%%%%%%%
%\input{background}
%%%%%%%%%%%%%%%%%%%%%%%%%%%%%%%%%%%%%%%%%%%%%%%%%%%%%%%
\section{Literature review}
\label{sec:litterature_review}
\subsection{Overview of smart cities and emergency response systems} 
\label{subsec:overview_smartcity_emergency}
A smart city is a collective endeavor leveraging communication and information technology to enhance sustainability, livability, and viability. Like any community, it inevitably encounters unforeseen emergencies demanding prompt attention to maintain normalcy. Yet, a sophisticated system is imperative to swiftly and efficiently address such emergent challenges \cite{emergency_system2019}. Numerous studies have delved into this domain. For instance, researchers in \cite{Hayat_2016} have delved into potential challenges confronting smart cities, advocating for a proactive disaster preparedness blueprint to fortify all aspects of smart city infrastructure against disasters and fire hazards. Similarly, \cite{Brandt_2016} has scrutinized various facets pertinent to urban dwellers in smart cities, including emergency responsiveness, while promoting strategic planning to fortify the resilience of smart cities. Additionally, \cite{ye_2016} has examined emergent systems in smart cities, focusing on real-time and distributed emergency resource management. They propose optimizing resource allocation using the Hierarchy Particle Swarm Optimization Algorithm (HPSO) to enhance resource quality and minimize costs and time. Moreover, they propose an emergency architecture comprising distinct layers: resource, network, sensing, smart service, and application. The research outlined in \cite{emergency_system2022} presents sensor-based detection of emergency, and highlighted a critical consideration regarding emergency detection is the level of reliability inherent in the entire process, which can be associated with various challenges. Firstly, sensor failures may occur, impeding the prompt identification of emergencies. Additionally, network failures may also arise, resulting in delays or outright prevention of emergency alert dissemination. In such instances, to enhance the expected quality of detection when utilizing sensor node networks, several strategies can be employed, including sensor redundancy, transmission reliability, and fault tolerance in sensor networks. Palmieri et al. \cite{PALMIERI2016810} propose a hybrid cloud architecture designed to oversee computing and storage resources essential for command and control operations during emergency situations. The architecture is supplemented by a pioneering first responder localization service that integrates signals from landmarks positioned by first responders at the crisis site with data gathered from motion sensors, offering a comprehensive positioning approach. In \cite{kijin2024@isASmart}, authors propose a systematic methodology centered on text mining to review several approaches on smart technologies during emergencies. Their research identified the following three themes: (a) emergency response, which encompasses emergency management, traffic and unmanned aerial vehicles, waste disposal, and contact tracing; (b) motivation and outcome, which includes sub-themes such as smart urbanism, quality of life, and the economy; and (c) technology and data, which covers social media, machine learning, the Internet of Things, data-driven applications, and object detection. Research in \cite{Elvas2021DisasterMI} aimed to address the need for Smart Cities to evaluate technologies and tools for disaster resilience, enabling them to wisely utilize resources and prioritize the most suitable options for their specific needs. The study identified 24 technologies and tools to create, sustain, and enhance resilience within Smart Cities. It emphasized the importance of collecting and managing citywide geodata and fostering public participation. The study identified four key factors for assessing these technologies: their impact on society, adoption speed by Smart Cities, maturity of the technology, and the capabilities offered to the community.

In the landscape of research pertaining to emergency response systems, there exists a notable gap in addressing the unique challenges posed by regions with limited infrastructure, exemplified by environments akin to that of Chad. As illustrated by the few presented studies above, previous studies have predominantly favored architectures reliant on cloud-based infrastructure or have operated under the assumption of uninterrupted network connectivity, overlooking the realities of resource-constrained settings. Furthermore, while incident response time is a critical factor in emergency scenarios, existing literature has largely neglected to explore methodologies for optimizing response times through the refinement of shortest path algorithms and the optimization of hardware resources responsible for executing these computations at the same time. This paper's research is thus positioned to bridge these gaps by proposing an innovative solution tackling the specific demands of resource-constrained environments as first-class citizen. By developing strategies that leverage localized computing resources and prioritize efficient algorithms, we aim to enhance the effectiveness and timeliness of emergency response operations in contexts where traditional approaches fall short.

\subsection{Field Programmable Gate Arrays (FPGAs) computing capabilities} 
\label{subsec:fpga_in_computing}
A Field Programmable Gate Array (FPGA) is a type of semiconductor device that is capable of being programmed to execute various digital functions across a broad spectrum \cite{wang2022review}. It comprises an array of programmable logic blocks and interconnections, allowing for the configuration to create personalized digital circuits \cite{siracusa2021comprehensive}. Unlike micro-controllers, which have fixed hardware and execute instructions sequentially, FPGAs can be reprogrammed to perform various tasks by modifying their internal connections and logic \cite{li2020survey}. The core functionality of FPGA technology is rooted in adaptive hardware, possessing the distinctive capability of being modified after manufacturing \cite{wang2022review}. Arrays of configurable hardware blocks can be interconnected as required, enabling the creation of highly efficient, domain-specific architectures for any application \cite{li2020survey}. FPGAs enable the parallel processing of numerous tasks, which proves particularly advantageous and efficient when dealing with extensive datasets. Sorting, filtering, and decimation operations are executed more effectively, freeing up primary processors to concentrate on signal processing algorithms \cite{boutros2021fpga}. In recent years, a number of studies like the ones of Benaicha et al. \cite{Benaicha2013} and Salim et al. \cite{Salim2020} have shown how it is possible to replace conventional processors with FPGA boards in order to increase computing speed. FPGAs are known for their ability to execute operations in parallel \cite{Chouchene2018,Roland2016}, which can be exploited in path finding and Dijkstra's algorithm. Ivan Fernandez et al. \cite{Fernandez2008} proposed the computation of Dijkstra algorithm for each node of a graph in order to get the minimum distance from each node to any other node in that graph. Assuming that the graph has $n$ nodes, their proposed FPGA architecture considers $n$ node processors; each node processor has 4 memories where it can read data concerning the graph and will compute the Dijkstra's algorithm for each node. However, this solution becomes useless when the graph has a large number of nodes and the usage of memory inside the FPGA component is not good in terms of scalability and execution speed. Guoqing et al \cite{Guoqing2016} proposed an architecture with the same idea; meaning the computation of Dijkstra algorithm for each node of the graph in order to get the minimum distance from each node to any other node in the graph. However, their solution is built for large graphs with a lot of nodes. Authors use priority queue to store all the nodes, and the Graph Processing Engines (GPEs) to compute the Dijkstra algorithm for each node. Let $n$ be the number of available GPEs. At each iteration, authors use the Dijkstra management module to remove $n$ nodes in the priority queue and send each node to a distinct GPE, which compute Dijkstra algorithm with the received node as source node. Unfortunately, authors did not integrate cash memory and every time, the Dijkstra algorithm must be computed for all the nodes. In \cite{Yanping2021}, Yanping et al. propose a solution for evacuating a building using the shortest path in the event of a fire. They propose an architecture where, In the event of a fire in a building, the IoT nodes send information to the edge server (in the building) about the location of the fire. The edge server runs Dijkstra's algorithm on its FPGA board, taking the location of the fire as the source of the graph in order to indicate the nearest outputs. Unfortunately, Dijkstra's algorithm is executed sequentially, so the authors do not take advantage of the parallelism offered by FPGA boards. This results in a slow execution of the Dijkstra algorithm when the number of nodes is very high. Mengqing et al. proposed in \cite{Mengqing2021} a solution to parallelize Dijkstra algorithm using several processors. The idea is to split the adjacency matrix representing the graph into $p$ sub-matrices ($p$ is the number of processors); and consider the same node $u$ as the source for each sub-matrix for each processor. At the end of each iteration, the resulting nodes $V_1$, $V_2$,…, $V_p$ are compared and the $V_i$ with the minimal distance is considered and given to the $p$ processors for the next iteration. In \cite{electronics13112167} authors introduce an optimization approach for calculating the shortest path in mobile robot route planning, specifically targeting real-time processing requirements with a high-performance solution. The approach utilizes an architecture embedded in dedicated hardware that emphasizes parallelism. By improving parallel exploration techniques, their solution not only boosts performance but also dynamically adapts to graph changes, accommodating random edge insertions or deletions as environmental conditions fluctuate. They present the developed architecture and its results, demonstrating efficient updates of obstacle matrices, resulting in a remarkable 120-fold improvement for 1024-node graphs. 

The presented FPGA-based works collectively illustrate the significant potential of FPGA technology in optimizing computational tasks, particularly for Dijkstra's algorithm. These works have demonstrated impressive improvements in computing speed through the parallel processing capabilities of FPGAs. However, despite these advancements, several limitations remain. Many of these solutions struggle with scalability and memory usage, making them less practical for large graphs. Additionally, the lack of integrated cache memory necessitates repeated computations, which undermines efficiency. Sequential execution in some approaches fails to fully exploit the inherent parallelism of FPGAs, resulting in slower performance when dealing with extensive node networks. Furthermore, these studies generally overlook the benefits of offline computing through cache memory integration and the advantages of edge computing. The absence of edge computing in these approaches means they do not leverage decentralized resources to enhance responsiveness and reduce reliance on continuous network connectivity. Addressing these gaps by integrating FPGA capabilities with edge computing and cache memory is particularly compelling for Chadian-like contexts, where infrastructure limitations and unreliable network connectivity demand innovative and resilient solutions. This integration could provide more robust, efficient, and scalable emergency response systems tailored to the unique challenges of such environments.

\subsection{Edge computing in smart cities} 
\label{subsec:EC_in_smart_cities}
Edge computing is an emerging computing paradigm which refers to a range of networks and devices at or near the user \cite{Cui2022Anovel}. Edge is about processing data closer to where it’s being generated, enabling processing at greater speeds and volumes, leading to greater action-led results in real time \cite{mcenroe2022survey}. With edge computing, it is possible to minimise the bandwidth requirements between IoT objects and data processing centres by undertaking analyses as close as possible to the data sources, and it also avoids the transmission of large amounts of irrelevant data to data centres or the cloud, bringing fluidity and speed of reaction. Edge computing increases the reliability and resilience of smart city systems. By distributing computing resources to edge devices, critical services can continue to operate locally in the event of network disruptions or cloud outages \cite{abdellatif2021medge}. Efficient disaster response relies on technology that facilitates prompt collection and analysis of data from various disaster sites within the affected region. For example, \cite{ROSAYYAN2023100697} proposed an emergency priority system to save live by giving right of way green signal to emergency vehicles such as ambulances and fire engines. The paper proposes an optimal control strategy for Emergency Vehicle Priority (EVP) using edge computing and IoT sensors for smart cities. The experiment utilized a GPS-based IoT sensor that continuously sends Location Information (LI) to the edge server. The edge server calculates the optimal timings using the proposed control strategy algorithm to clear the path for emergency vehicles.

The work in this paper leverages the advantages of edge computing to address critical limitations observed in existing FPGA-based solutions for emergency response optimization. By integrating edge computing, we decentralize computational resources, enabling real-time processing and reducing dependency on continuous network connectivity. Edge computing allows for localized data processing, ensuring that critical computations, such as the execution of Dijkstra's algorithm for route planning, can be performed efficiently and promptly at the edge of the network. This reduces latency and enhances the overall responsiveness of the system. Furthermore, the proposed architecture incorporates cache memory to store and reuse computed paths, thereby minimizing redundant calculations and improving efficiency.

%%%%%%%%%%%%%%%%%%%%%%%%%%%%%%%%%%%%%%%%%%%%%%%%%%%%%%%
\section{Proposed system architecture}
\label{sec:contribution}
In the realm of urban emergency response systems, the imperative for swift, efficient, and tailored solutions has become more pressing, especially within the dynamic landscape of smart cities. Recognizing the pivotal role of technology in meeting these demands, this paper's contribution focuses on the development of a novel system architecture that addresses the core challenges inherent in emergency response, particularly within Chadian-like contexts.
The architecture is designed with a keen emphasis on three fundamental issues: \emph{time efficiency}, \emph{resource allocation} and \emph{usability in Chadian-like contexts}.

\noindent\textbf{Time efficiency}:
Acknowledging the criticality of rapid response times in emergency situations, the proposed architecture prioritizes streamlined processes and optimized workflows to minimize delays and enhance overall response efficacy.

\noindent\textbf{Resource allocation}:
Efficient distribution of resources is paramount for effective emergency response. The provided architecture incorporates sophisticated algorithms and decentralized computing resources to ensure the judicious allocation of personnel, equipment, and supplies, thus maximizing the impact of available resources.

\noindent\textbf{Usability in Chadian-like contexts}:
Understanding the infrastructural, geographical, and socioeconomic challenges present in Chad, the architecture is crafted to be adaptable, intuitive, and responsive to the specific needs and conditions of Chadian-like environments. Indeed, designing an emergency response architecture tailored to Chadian-like contexts necessitates consideration of several key features to ensure effectiveness and adaptability. Some of these features include:
\begin{itemize}[itemsep=-0.3em]
    \item \textit{Offline capabilities}: Given the intermittent or unreliable connectivity often encountered in remote or underdeveloped areas, the architecture should support offline operation of essential functionalities. This ensures that emergency response operations can continue even in the absence of a stable network connection.

    \item \textit{Fault tolerance}: The system should be resilient to failures or disruptions, whether due to power outages, equipment malfunctions, or network issues.

    \item \textit{Energy efficiency}: Given the prevalence of energy scarcity in some areas, the system should prioritize energy-efficient operation.

    \item \textit{Localized decision-making}: Empowering local emergency responders with decision-making capabilities can enhance response agility and effectiveness.
    
    \item \textit{Adaptive response protocols}: Establishing flexible response protocols that allow for rapid adjustment based on real-time feedback and emerging challenges is essential. The architecture should support adaptive decision-making processes that enable responders to improvise and innovate in response to unexpected obstacles or constraints.

    \item \textit{Low bandwidth requirements}: To accommodate limited network bandwidth in certain areas, the architecture should minimize data transfer requirements and prioritize essential communication.
\end{itemize}

\subsection{Overview of the proposed architecture}
\label{sec:proposed_architecture}
Considering the context of smart cities, we propose a solution to speed up the response of a city's emergency services (police, hospitals, fire brigade, etc.) in the event of an incident in a building. We consider that there is several edge servers in the city, and each edge server is equipped with a FPGA-based board. In each city's building, several IoT objects equipped with diverse sensing capabilities are deployed. In the event of an incident, these send the information directly to the nearest edge server. 

The edge server, which knows the city and represents it in the form of a graph, will consider the location of the incident as the source node and will run the Dijkstra algorithm on its FPGA board. Running this algorithm produces the minimum path to the incident scene for each emergency service. The edge server sends this result to the various services in order of priority, with the centre closest to the incident scene having the highest priority for each service.

Each emergency service has a system that receives alerts from edge servers. When the service with the highest priority receives the alert from the edge server, it takes the necessary steps to send the vehicles with all the necessary information. If this service is unable to react, it can notify the edge server so that the latter can notify the service with the highest priority among the remaining services, or it can notify this service directly if the connection to the edge server is interrupted.

On the way to the incident scene, the vehicles follow the path given to them by the processing system located on the emergency services premises. If there are any unforeseen circumstances on the road and the route has to be changed, the vehicle can notify the edge servers for a rapid recalculation of a new route, or it can recalculate the route itself, since it is also equipped with FPGA-based devices. 

\begin{figure}[h!]
    \centering
    \includegraphics[width=1.0\linewidth]{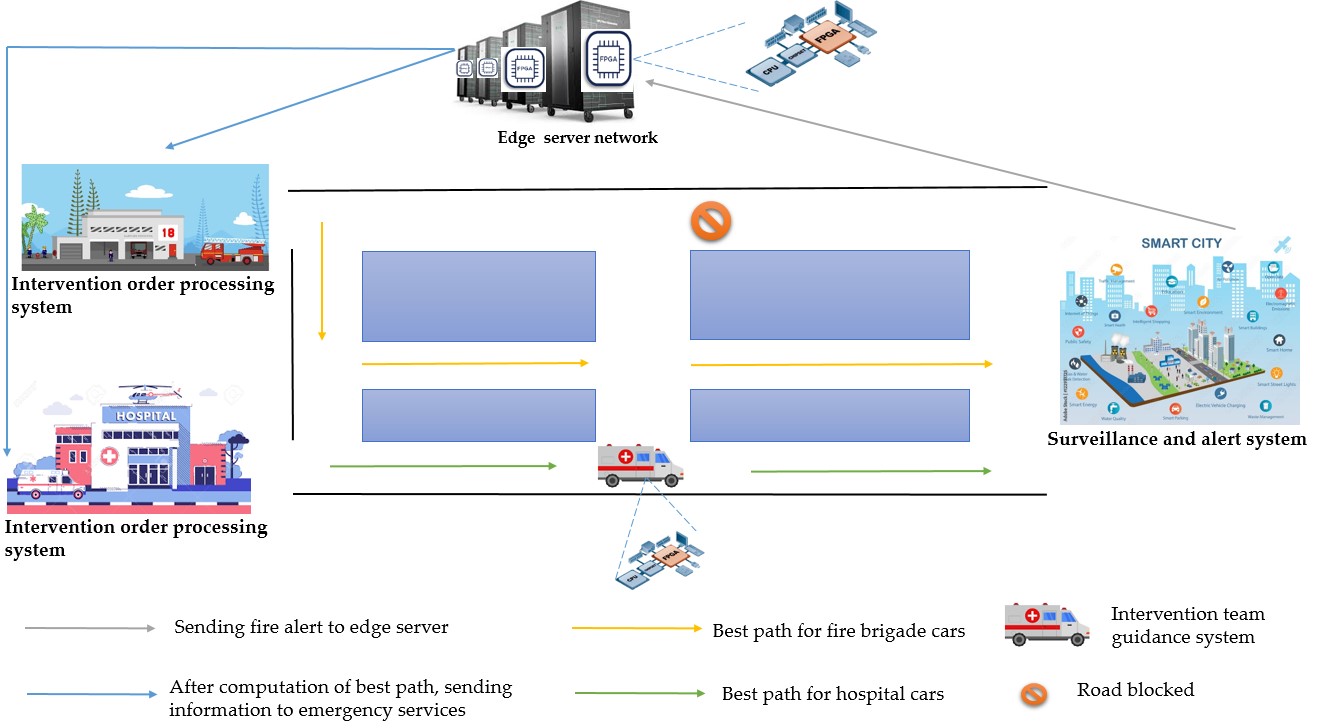}
    \caption{\label{fig:overview_proposed_architecture} Overview of the proposed architecture.}
\end{figure}

Concretely, the proposed architecture comprises several interconnected subsystems (see Figure \ref{fig:overview_proposed_architecture}):
\begin{enumerate}[itemsep=-0.3em]
    \item \textbf{The Surveillance and Alerting System (SAS)}: This subsystem consists of a network of IoT devices strategically deployed in a city building or area to detect incident and alert authorities of emergencies. These devices, equipped with various sensors, can detect incidents such as fires, security threats, or natural disasters and trigger alerts to initiate response actions.

    \item \textbf{The Edge Server Network (ESN)}: The architecture includes a cluster of interconnected edge servers, each equipped with high computational capabilities and FPGA-based systems. These servers receive alerts from the surveillance system's main IoT device and perform distributed computations, including route optimization using a distributed version of Dijkstra's algorithm.

    \item \textbf{The Intervention Order Processing System (IOPS)}: Installed within each intervention service, this system receives alerts from edge servers and determines the appropriate response based on priority. It prepares intervention teams for immediate action, coordinates communication with the edge servers, and provides guidance to intervention teams during operations.

    \item \textbf{The Intervention Team Guidance System (ITGS)}: This system, based on FPGA technology, provides real-time guidance to intervention teams, displaying optimized routes and dynamically recalculating paths if necessary. It communicates with the intervention order processing system and the edge servers for continuous monitoring and decision support.
\end{enumerate}

\subsection{The Surveillance and Alerting System (SAS)}
The SAS is a vital component of this architecture, comprising a network of interconnected IoT devices equipped with diverse sensing capabilities to detect and signal incidents occurring at specific locations, thus initiating an intervention. Strategically deployed in key areas of the location, each device is tailored to specific needs, such as fire monitoring or detection of terrorist threats, etc. To avoid sending several messages containing the same information directly to the edge server, the network each SAS has a special FPGA-based IoT device named central alert system, whose role is to collect the information about the place of the incident, and trigger only one request to the edge servers. By this way, the edge servers will process only one request for one incident, instead of processing several requests for the same incident. Extensive literature exists detailing the construction of systems behaving like the SAS, providing a robust foundation for implementation. This is an algorithmic description of the SAS behaviour:
\begin{enumerate}[itemsep=-0.3em]
    \item \textbf{Initialization}:
        Each IoT device is configured with predefined monitoring parameters based on its designated function and location within the urban environment.

    \item \textbf{Sensing and detection}:
        At regular intervals, the IoT devices collect sensor data relevant to their designated monitoring tasks. Data processing algorithms analyze collected data to detect anomalies indicative of emergency incidents.

    \item \textbf{Incident identification}:
        Upon detecting a potential emergency incident, the IoT device evaluates the severity and nature of the situation. If the incident meets predefined criteria for triggering an alert, the device proceeds to alert generation.

    \item \textbf{Alert generation}:
        The IoT device generates an alert message containing detailed information about the incident, including its location and characteristics. The alert message is transmitted to the central Alerting System for further processing and dissemination.

    \item \textbf{Central alerting system}:
        Upon receiving alert messages, the central alerting system analyzes the data to verify the occurrence of an incident. If confirmed, the system selects an edge server from a preconfigured list based on proximity and availability.

    \item \textbf{Edge server selection}:
        The Central Alerting System iterates through the list of edge servers, pinging each one until a response is received. The first responding server is designated to receive the alert message. If the connection is interrupted during transmission, the system retries with another available server.

    \item \textbf{Alert transmission}:
        The selected edge server receives the alert message and prepares it for dissemination to relevant emergency response teams and authorities. The alert contains comprehensive information about the incident, enabling responders to assess the situation and initiate appropriate actions.
\end{enumerate}
\begin{figure}[h!]
    \centering
    \includegraphics[scale=0.15]{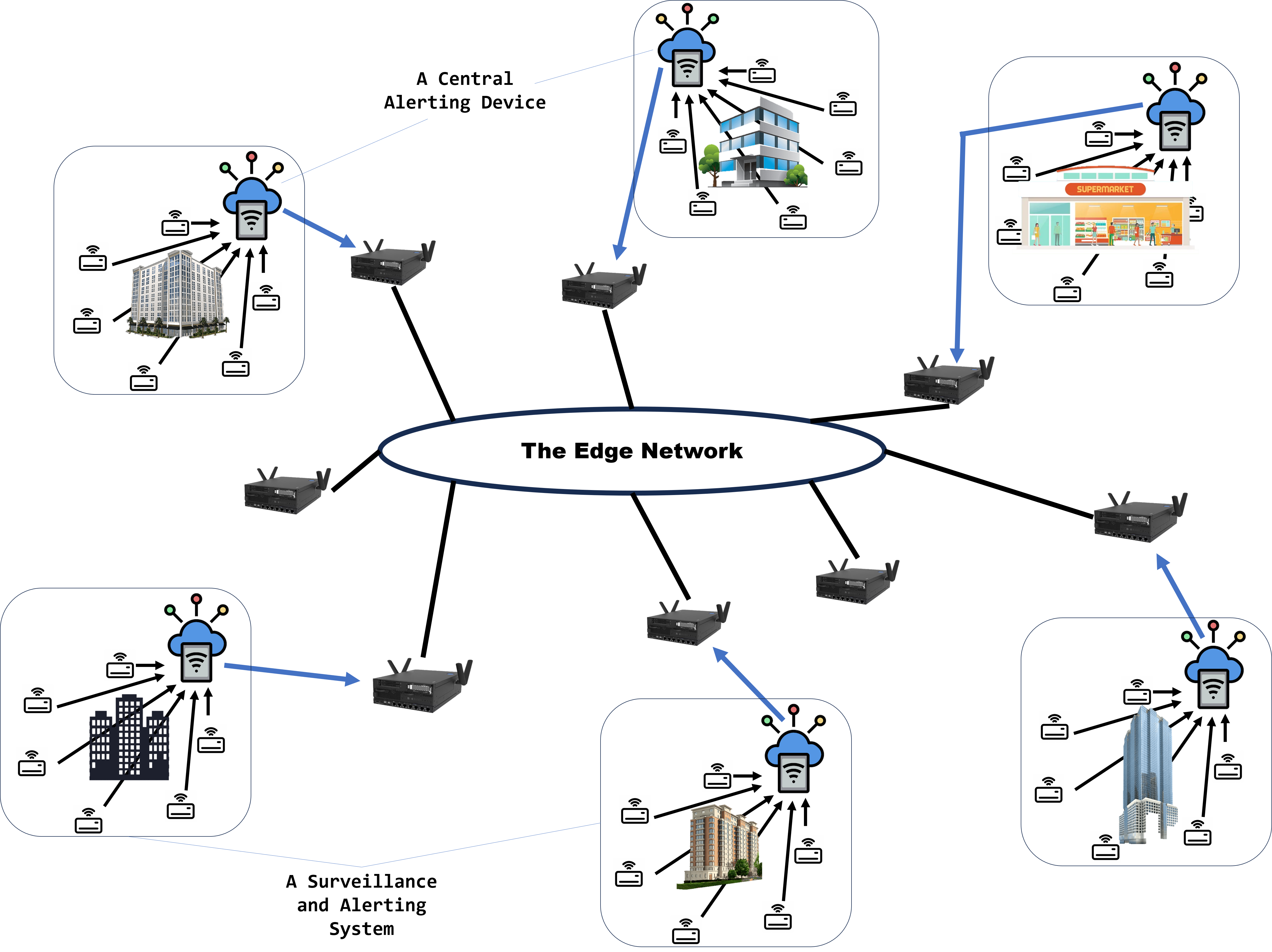}
    \caption{\label{fig:overview_sas_edge} Overview of the edge network.}
\end{figure}
Figure \ref{fig:overview_sas_edge} gives an overview of the network architecture and highlights the communication between the IoT devices within a SAS and between the SAS and the edge server network. In it, one can see that a SAS deployed to monitor a particular area of interest is made up of several IoT devices forming a local network. Some of these devices are responsible for collecting data (hazards) and sending them to the central alerting system (a special device designed using FPGA technology) so that it can analyse them to detect potential incidents. When an incident is detected, the details are sent by the central alerting system to the nearest active edge server to trigger the intervention process.

\subsection{The Edge Server Network (ESN)}
\subsubsection{The ESN behaviour}
The Edge Server Network (ESN) serves as a pivotal component within the proposed architecture, facilitating efficient and decentralized emergency response operations. Comprising interconnected servers forming a cluster, each ESN server boasts substantial computational and storage capabilities, enabling distributed task execution. Equipped with FPGA-based systems specialized in shortest path calculation using a distributed version of the Dijkstra algorithm, these servers play a crucial role in determining optimal routes for intervention teams.

Upon receiving alerts, each ESN server initiates the computation of the shortest paths from intervention services to the distressed location. Leveraging its onboard FPGA device, the server orchestrates the distributed execution of the Dijkstra algorithm, utilizing its access to the smart city representation. This representation encompasses surveillance points, intervention services, and intermediary landmarks, forming a comprehensive graph of traversable paths within the city.

Furthermore, each ESN server maintains a cache memory storing previously evaluated shortest paths to monitored locations. This caching mechanism minimizes redundant path calculations, updating the cache periodically to synchronize with any changes in the city's infrastructure. By leveraging cached data, ESN servers optimize computational efficiency, recalculating paths only when necessary.

Post-path computation, the server prioritizes intervention services based on the severity and nature of the incident, generating an ordered list of response teams. It then dispatches alerts to the identified services, providing essential intervention details such as the shortest path and threat characteristics. The server awaits confirmation from the highest-priority service before proceeding with subsequent interventions. In the event of non-confirmation or declination, the server redirects the alert to the next prioritized service, ensuring swift and decisive response actions.

Throughout the intervention process, the server maintains communication with deployed intervention teams, facilitating resource allocation and real-time coordination. In response to resource requests from active intervention sites, the server identifies and dispatches the nearest and most suitable services, ensuring optimal resource utilization.

Ultimately, the intervention concludes upon receipt of confirmation messages from all active intervention sites, prompting the server to disseminate intervention completion notifications to all alerted sites.

\subsubsection{The physical architecture of an edge server in the system}
\begin{figure*}[h!]
    \centering
    \includegraphics[width=1.0\linewidth]{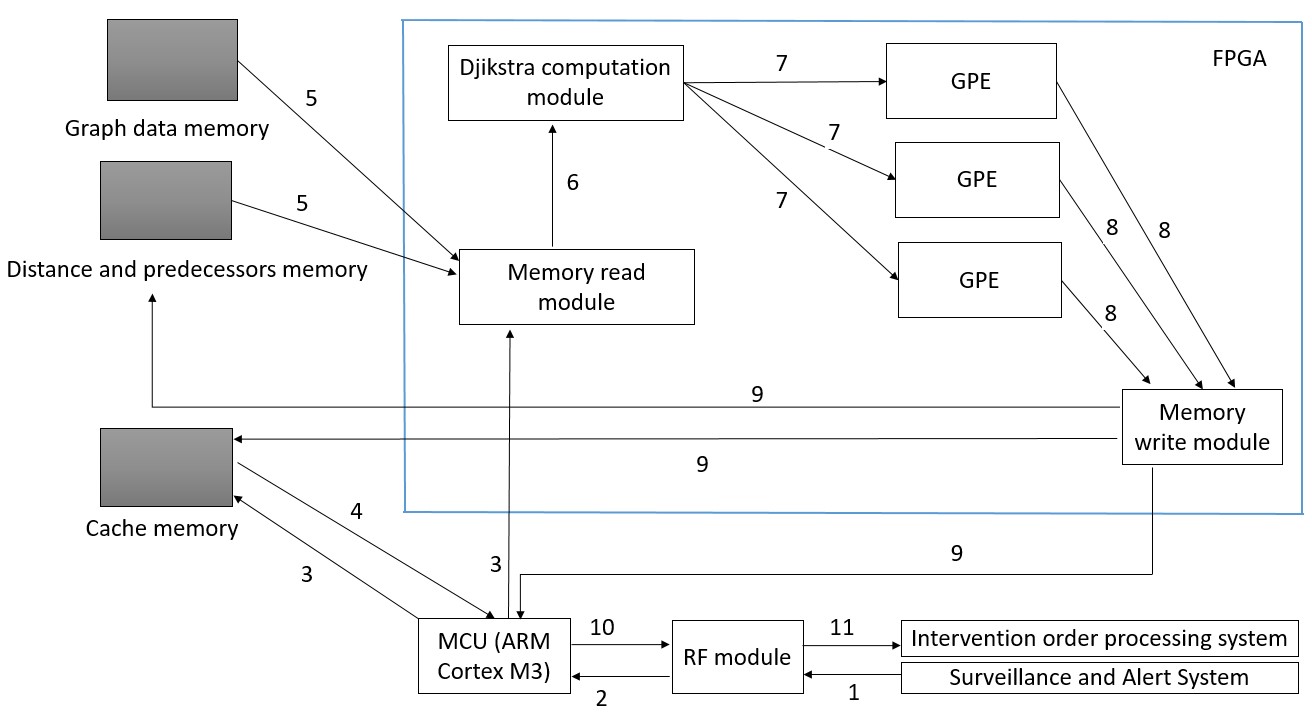}
    \caption{\label{fig:architecture_modele} The physical architecture of an edge server in the system.}
\end{figure*}

Figure \ref{fig:architecture_modele} describes the core components of an edge server in the system and how they behave, from the reception of a request on the radio frequency module, its processing by the FPGA module, until the sending of the alert to the emergency services, while saving the obtained result in the cache memory.

The edge server architecture in Figure \ref{fig:architecture_modele} is composed by the following:
\begin{itemize}[itemsep=-0.3em]
    \item \textbf{The RF module}: the radio frequency communication module of the edge server that is responsible of receiving alerts from the SAS, and send information to the Intervention Order Processing System installed on the premises of emergency services;
    \item \textbf{The MCU (Micro-Controller Unit)}: processor responsible of the computation of simple operations;
    \item \textbf{The graph and data memory}: stores the representation of the smart city as a graph;
    \item \textbf{The distance and predecessors memory}: stores for a node u, its predecessor in the graph, and the minimum distance from source node to node u;
    \item \textbf{The cache memory}: stores for a node u, the best paths found after the execution of Dijkstra algorithm. Instead of re-computing Dijkstra algorithm for a given node u that has already been used before, the content of this memory can be used in order to provide the best path quickly;
    \item \textbf{The FPGA module}: dedicated to the parallel execution of the Dijkstra algorithm according to the parameters supplied to it. The FPGA module is composed of:
    \begin{itemize}[itemsep=-0.3em]
        \item \textbf{The memory read/write module}: I/O module used to read and write data outside the FPGA component;
        \item \textbf{The Dijkstra computation module}: used to store all the data and instructions for Dijkstra algorithm computation;
        \item \textbf{Several Graph Processing Engines (GPE)}: a GPE is a calculation module in the FPGA.
    \end{itemize}
\end{itemize}

Given this physical architecture, the behaviour of an edge server in the system can be described using the following steps
\begin{enumerate}[itemsep=-0.3em]
    \item \textbf{Reception of alert}:
    After receiving and responding to the ping from a Surveillance and Alert System, the selected edge server receives the alert of an incident on its RF module. It identifies the location $u$ of the incident in the request that it receives.
    \item \textbf{Sending location of the incident}:
    The location of the incident, identified as a point $u$, is sent to the MCU by the RF Module for further operations.
    \item \textbf{Pre-processing and FPGA solicitation}:
    The MCU receives the location $u$ of the incident. It first checks if in the cache memory there is already computed path for location $u$. If it is not the case, it sends location $u$ to the memory read module of the FPGA board.
    \item \textbf{Providing existing path}:
    If the MCU find the cache memory already have a path for the location $u$, it directly notify the RF module to send the result to the emergency services.
    \item \textbf{Data initialization}:
    Data are initialized in the memory read module in order to prepare for the computation of the shortest paths to location $u$ using Dijkstra algorithm. Thus, the memory read module gets the graph structure from the Graph data memory, and the distance and predecessor of each node from the distance and predecessor’s memory.
    \item \textbf{Transmission of data}:
    Graph structure, source node $u$ and distance and predecessors are transferred to the Dijkstra Computation module. The latter splits the adjacency matrix obtained from the graph structure into $p$ sub-matrices ($p$ representing the number of available GPEs).
    \item \textbf{Dijkstra algorithm execution}:
    The given node $u$ (the location of the incident) is given as the entry point to all the GPE in charge of the execution of Dijkstra algorithm. Each GPE executes Algorithm \ref{alg:Dijkstra_algorithm} in parallel while considering each sub-matrix given by the Dijkstra computation module. The behaviour of this stage is explained at section \ref{subsec:execution_of_dijkstra_algo}.
    \item \textbf{Sending results to memory write module}:
    The obtained results by the different GPEs are synchronized, and a single shortest path is obtained for each emergency service to the location of incident $u$. Results of the execution and synchronization are sent to the memory write module.
    \item \textbf{Local memory updates and providing results}:
    The memory write module which is aware of the obtained results after the execution of Dijkstra algorithm for a given location $u$, sends that results in parallel to the MCU module, and the cache memory. It also updates the distance and predecessors memory with values obtained during the execution of the algorithm.
    \item \textbf{Sending results to RF module}:
    After receiving results from the memory write module, the MCU orders the RF module to broadcast the alert to the Intervention Order Processing Systems.
    \item \textbf{Sending alert to the Intervention Order Processing System}:
    The RF module which is in charge of the communication between an edge server and other devices, forwards the obtained results to the Intervention Order Processing Systems. These results contain the list of the emergency services with their priority starting from the emergency service with the highest priority to the service with the lowest one.
\end{enumerate}

\subsubsection{Parallel execution of Dijkstra algorithm} 
\label{subsec:execution_of_dijkstra_algo}
\begin{figure*}
    \centering
    \includegraphics[width=1\linewidth]{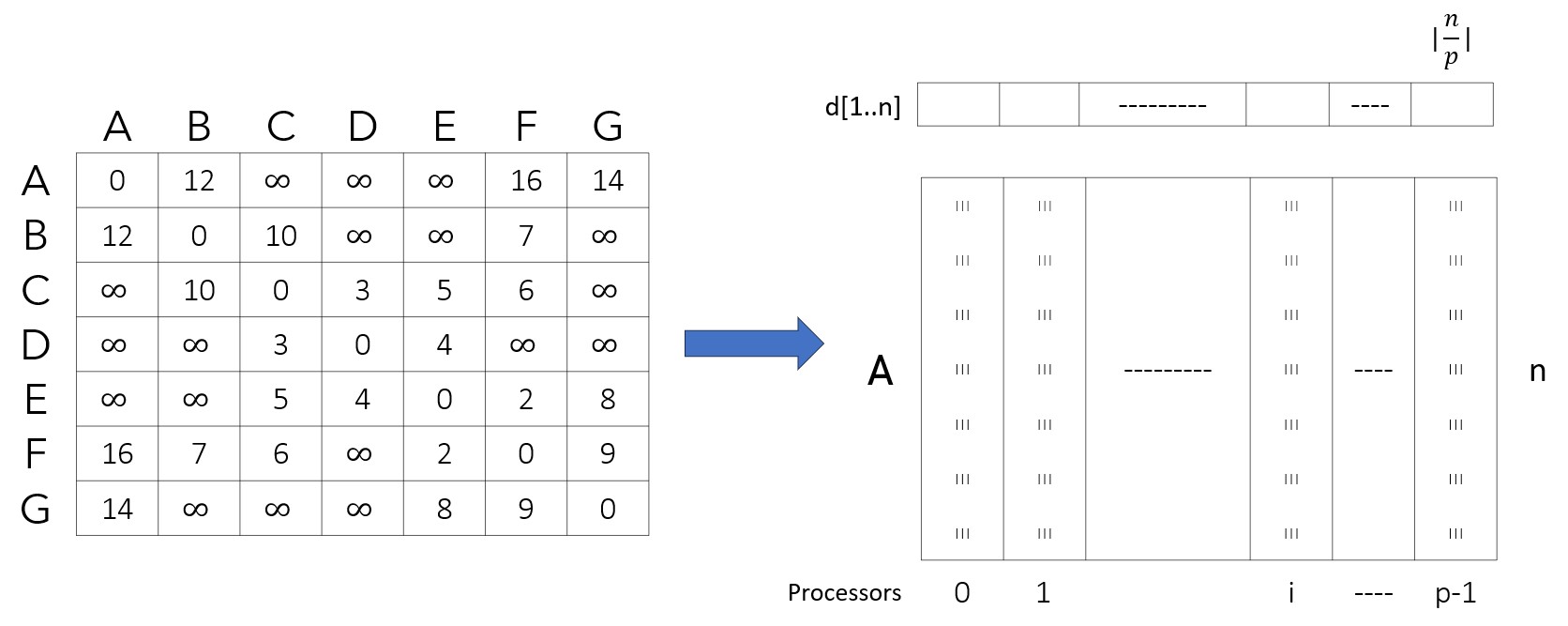}
    \caption{\label{fig:matrix_to_submatrices} Splitting adjacency matrix to sub-matrices.}
\end{figure*}

Remember that the edge server knows the city and has a graph representation of it in its memory. Two different data structures are mainly used to represent a graph: adjacency matrix and adjacency lists. In this case, the adjacency matrix is used to represent the  graph because it is efficient, easy to understand, easy to update and rapid in terms of operations speed. The adjacency matrix $M$ is a table representation of the relationships between nodes. Rows and columns of the matrix are labelled by nodes' IDs, and the values of cells indicate the distance between the nodes. The edge server divides the adjacency matrix representing the graph into $p$ sub-matrices (as illustrated on Figure \ref{fig:matrix_to_submatrices}). Each GPE in the FPGA board  will execute the Dijkstra algorithm for a given sub-matrix.

\RestyleAlgo{ruled}
\begin{algorithm}
    \LinesNumbered
    \caption{Dijkstra's Single Source Shortest Path Algorithm}\label{alg:Dijkstra_algorithm}
    \KwIn{$V$ \tcc{The set of vertices}}
    \myinput{$E$ \tcc{The set of edges}}
    \myinput{$w$ \tcc{The distance from one node to another}}
    \myinput{$s$ \tcc{The source vertex}}
    
    \KwResult{$d[v]$ \tcc{The shortest distance from $s$ to each vertex $v$}}

    $d[s] \leftarrow 0$ \tcc{Distance from $s$ to itself is 0}
    $S \leftarrow \{s\}$ \tcc{The set of vertices for which the shortest path has been found}
    $U \leftarrow V$ \tcc{Remaining vertices to be processed}
    \ForEach{$v \in U$}{
        \eIf{$(s,v) \in E$}{
            $d[v] \leftarrow w(s,v)$ \tcc{Distance from $s$ to adjacent vertices}
        }{
            $d[v] \leftarrow \infty$ \tcc{Infinite distance for non-adjacent vertices}
        }
    }

    \While{$U \neq \emptyset$}{
        $u \leftarrow$ vertex $v$ in $U$ with the smallest $d[v]$ \tcc{Select the vertex $u$ with the minimum distance}
        $S \leftarrow S \cup \{u\}$ \tcc{Add $u$ to the set of processed vertices}
        $U \leftarrow U \setminus \{u\}$ \tcc{Remove $u$ from the remaining vertices}
        \ForEach{$v \in \text{Adj}[u]$}{
            \tcc{Adj[u] refers to the set of vertices adjacent to $u$}
            \If{$d[v] > d[u] + w(u,v)$}{
                $d[v] \leftarrow d[u] + w(u,v)$ \tcc{Relax the edge and update the distance}
            }
        }
    }
\end{algorithm}

\begin{figure*}
    \centering
    \includegraphics[width=1\linewidth]{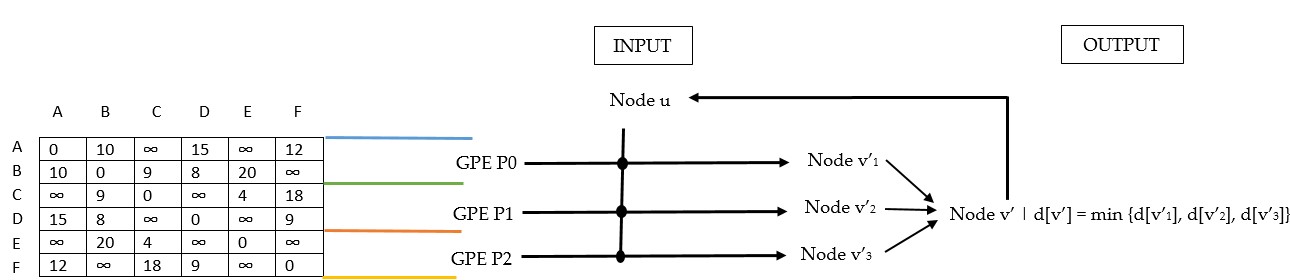}
    \caption{\label{fig:parralel_Dijkstra_scheme} Parallelism of Dijkstra algorithm using GPE.}
\end{figure*}

A sequential version of the Dijkstra algorithm for the shortest path computation is given in algorithm \ref{alg:Dijkstra_algorithm}. According to it, Dijkstra’s algorithm is iterative. After each iteration, a new vertex is added to the computed set $S$. Very often, to parallelize an algorithm, we try to run its iterative structures in parallel on several CPUs. In this case, Since the value of $d[v]$ for a vertex $v$ may change every time a new vertex $u$ is added in $S$, it is hard to select more than one vertex; and then it is not easy to perform different iterations of the while loop in parallel. To remedy this, we use the proposed architecture (Figure \ref{fig:architecture_modele}) to run this algorithm in parallel on the FPGA board. The following describes our approach:

\begin{itemize}[itemsep=-0.3em]
    \item The adjacency matrix representing the graph is subdivided into $p$ sub-matrices;
    \item let $n$ be the number of vertices. Each sub-matrix has $\frac{n}{p}$ consecutive vertices, and the work associated with each sub-matrix is assigned to a different GPE;
    \item Let $V_{i}$ be the subset of vertices assigned to GPE $P_{i}$ for $i = 0, 1,\cdots, p - 1$. Each GPE $P_{i}$ stores the part of the array $d$ that corresponds to $V_{i}$;
    \item Each GPE $P_{i}$ computes $d_i[u] = min\left\{d_i[v] ~|~ v \in (V-S)\bigcap V_i\right\}$ during each iteration of the while loop. The obtained minimums are sent to GPE $P_0$ so that, it finds the global minimum ;
    \item GPE $P_{0}$ now holds the closest vertex $u$ to the source. GPE $P_{0}$ broadcasts $u$ to all GPEs so that the GPE $P_{i}$ responsible for vertex $u$, marks $u$ as belonging to set $S$. Finally, each GPEs updates the values of $d[v]$ for its local vertices;
    \item When a new vertex $u$ is inserted into $S$, the values of $d[v]$ for $v \in (V-S)$ must be updated. The GPE responsible for $v$ must know the weight of the edge $(u,v)$. Hence, each GPE $P_{i}$ needs to store the columns of the weighted adjacency matrix corresponding to set $S$ of the vertices assigned to it. 
\end{itemize}
This process is described on Figure \ref{fig:parralel_Dijkstra_scheme}; a pseudo-code explaining the proposed approach is also provided after this figure.

\begin{tcolorbox}[title=Concurrent Dijkstra's Algorithm, colback=white, colframe=black, boxrule=0.8mm, sharp corners]
\noindent\textbf{Input:}
\begin{itemize}[itemsep=-0.3em]
    \item $V$: The set of vertices
    \item $E$: The set of edges
    \item $p$: The number of GPEs (processors)
    \item $s$: The source vertex
    \item $w(u,v)$: The weight of edge $(u,v)$
\end{itemize}

\textbf{Output:} The array $d$ containing the shortest path distances from the source vertex to all other vertices.

\textbf{Initialization:}
\begin{itemize}[itemsep=-0.3em]
    \item Divide the adjacency matrix into $p$ sub-matrices, one per GPE;
    \item Assign each sub-matrix to GPE $P_i$, where $i = 0, 1, \dots, p-1$;
    \item Each GPE $P_i$ stores the corresponding part of the distance array $d$ for vertices in $V_i$;
    \item Set $S \gets \emptyset$, where $S$ is the set of vertices for which the shortest path has been found;
    \item For each vertex $v$ in $V_i$:
    \begin{itemize}[itemsep=-0.3em]
        \item If $v$ is the source vertex, set $d_i[v] \gets 0$;
        \item Otherwise, set $d_i[v] \gets w(s,v)$.
    \end{itemize}
\end{itemize}

\textbf{Main Loop on each GPE (While $S \neq V$):}
\begin{itemize}[itemsep=-0.3em]
    \item Each GPE $P_i$ finds the vertex $u_i$ with the smallest distance value among the vertices in its subset $V_i$ that are not yet included in $S$;
    \item The $\{u_i \}$ are sent to GPE $P_0$ for the determination of the global minimum $u$;
    \item $u$ is sent to all GPEs by GPE $P_0$;
    \item Each GPE $P_i$ marks $u$ as belonging to $S$;
    \item For each $v \in (V - S) \cap V_i$: If $d_i[v] > d[u] + w(u,v)$, update $d_i[v] \gets d[u] + w(u,v)$.
\end{itemize}
\end{tcolorbox}

\subsection{The Intervention Order Processing System (IOPS)}
The IOPS serves as a critical component installed within each intervention service, enabling reception and processing of messages originating from edge servers. Upon receiving an alert, this system evaluates whether immediate intervention is warranted based on its priority ranking. If immediate action is not required, the service remains on standby, prepared to intervene if necessary. In cases where immediate intervention is deemed necessary, the system transfers pertinent information, including the route to the incident location, the coordinating edge server's address, and the city map graph received from the server, to a mobile guidance system carried by the intervention team. The Intervention Order Processing System may be connected to various alert services (such as email, SMS, or sirens) to promptly notify agents upon receiving an intervention alert.
\begin{enumerate}[itemsep=-0.3em]
    \item \textbf{Reception of alert}:
        The IOPS receives alert messages from edge servers, containing information about the incident and its location.

    \item \textbf{Priority evaluation}:
        Based on predefined priority criteria, the system assesses the urgency of the alert and determines whether immediate intervention is required.

    \item \textbf{Standby or immediate action}:
        If immediate intervention is not necessary, the service remains on standby, ready to respond if the situation escalates. In the case of an urgent alert requiring immediate action, the system proceeds to coordinate intervention efforts.

    \item \textbf{Information transfer}:
        The system transfers essential information, including the incident location, route guidance, and coordinating server details, to a mobile guidance system carried by the intervention team.

    \item \textbf{Integration with alert services}:
        The system interfaces with various alert services, such as email, SMS, or sirens, to promptly notify intervention agents upon receiving an alert.
\end{enumerate}

\subsection{The Intervention Team Guidance System (ITGS)}
The ITGS represents an essential component of the architecture, leveraging FPGA-based technology to provide real-time guidance and support to intervention teams during emergency response operations. Equipped with a display screen, this system visualizes the optimal route for intervention teams to follow. Moreover, it possesses computational capabilities to evaluate alternative routes in case the initially proposed path is obstructed or communication with the server is disrupted. This adaptability ensures that intervention teams can swiftly overcome obstacles and continue their mission uninterrupted. Additionally, the system facilitates communication with the IOPS and the central server throughout the operation, enabling continuous monitoring and decision support. It behaves like described below:
\begin{enumerate}[itemsep=-0.3em]
    \item \textbf{Initialization}:
        The ITGS initializes upon receiving instructions from the IOPS, including the incident location and route guidance.

    \item \textbf{Route representation}:
        The system displays the recommended route on its screen, providing intervention teams with visual guidance to the incident location.

    \item \textbf{Real-time evaluation}:
        As intervention teams progress towards the destination, the system continuously evaluates their position and the feasibility of the proposed route.

    \item \textbf{Alternative route calculation}:
        If obstacles are encountered or communication with the server is disrupted, the system utilizes its computational capabilities to assess alternative routes.
        Using local algorithms, the system determines the most efficient alternative path based on available information.

    \item \textbf{Dynamic route adjustment}:
        The system dynamically adjusts the displayed route in real-time, guiding intervention teams along the newly calculated path if necessary.

    \item \textbf{Communication with processing system and server}:
        Throughout the intervention, the system maintains communication with the IOPS and the central server. This communication facilitates continuous monitoring of the intervention progress and enables timely decision-making support from the central command.
\end{enumerate}

The ITGS plays a crucial role in ensuring the effectiveness and adaptability of intervention teams during emergency response operations, empowering them to navigate complex environments and overcome challenges encountered in Chad-like contexts.

\subsection{Discussion and perspectives}
In addressing the critical challenges of time efficiency, resource allocation, and usability in Chadian-like contexts, the proposed architecture presents a comprehensive solution that integrates multiple components to optimize emergency response processes.

\noindent\textbf{Time efficiency}:
The architecture's design prioritizes swift response times through the coordinated functioning of its constituent components. The SAS swiftly detects incidents, triggering immediate alerts to initiate response actions. This rapid detection is complemented by the ESN's distributed computing capabilities, which expedite the calculation of optimal paths to incident sites. Additionally, the IOPS streamlines response tasks, ensuring that resources are allocated promptly and effectively. The ITGS further enhances time efficiency by providing real-time guidance to intervention teams, minimizing delays in reaching incident sites.

\noindent\textbf{Resource allocation}:
Efficient resource allocation is facilitated by the architecture's intelligent coordination mechanisms. While the SAS detects incidents, its primary contribution lies in initiating the response process rather than directly allocating resources. In contrast, the ESN optimizes resource deployment by calculating optimal paths to incident sites, ensuring that personnel and equipment are directed to critical areas in a timely manner. The IOPS plays a central role in resource allocation by prioritizing response tasks based on the severity and urgency of each incident, thus optimizing resource utilization. However, the ITGS focuses on guiding intervention teams rather than directly allocating resources.

\noindent\textbf{Usability in Chadian-like contexts}:
Adaptability and usability in challenging environments are key considerations in the architecture's design. The SAS enhances usability by providing continuous incident monitoring, particularly in areas with intermittent connectivity. Similarly, the ESN's fault tolerance and energy efficiency features make it well-suited for deployment in resource-constrained environments, ensuring adaptability to infrastructure constraints. However, the usability of the IOPS may be limited by the availability of reliable communication infrastructure in Chadian-like contexts. Despite these challenges, the ITGS enhances usability by providing real-time guidance to intervention teams, even in areas with limited infrastructure.

\subsubsection{Strengths and capabilities}
\noindent\textbf{Enhanced Emergency Response Efficiency}: The architecture leverages advanced technologies such as IoT devices, edge servers, and FPGA-based systems to significantly enhance the efficiency and effectiveness of emergency response operations. Real-time data collection and analysis enable rapid detection of incidents, while distributed computation facilitates optimized route planning and coordination of intervention teams.
    
\noindent\textbf{Adaptability and flexibility}: The distributed nature of the architecture, coupled with its computational capabilities, ensures adaptability and flexibility in responding to dynamic urban environments. The ability to dynamically adjust routes and response strategies based on real-time data and changing conditions enhances the agility and responsiveness of intervention teams.

\noindent\textbf{Integration of cutting-edge technologies}: By integrating technologies such as FPGA-based computation, edge computing, and IoT sensors, the architecture represents a state-of-the-art solution for addressing the complex challenges of emergency response in Chad-like environments. These technologies offer unprecedented capabilities for rapid decision-making, precise navigation, and seamless communication among responders.

\subsubsection{Drawbacks and challenges}
\noindent\textbf{Theoretical nature of the proposal}: While the paper outlines a comprehensive architectural design for enhancing emergency response efficiency, it's important to acknowledge that the presented concepts remain theoretical. The practical implementation of such a complex system entails significant challenges and requires extensive research, development, and testing efforts. Transitioning from theoretical concepts to real-world deployment involves addressing practical constraints, technical limitations, and operational considerations, highlighting the need for further work to translate the proposed architecture into tangible solutions.
    
\noindent\textbf{Scalability and deployment complexity}: Deploying and scaling the architecture across a city-wide network may pose significant logistical and technical challenges. Coordinating the installation of IoT devices, edge servers, and communication infrastructure, as well as ensuring interoperability with existing systems, requires careful planning and resource allocation.

\noindent\textbf{Reliability and resilience concerns}: The reliance on interconnected devices and communication networks introduces potential points of failure, raising concerns about reliability and resilience. Interruptions in connectivity or malfunctions in hardware could compromise the effectiveness of emergency response operations, highlighting the need for robust redundancy measures and disaster recovery protocols.

\noindent\textbf{Data security and privacy risks}: The collection, processing, and transmission of sensitive data from IoT devices raise concerns about data security and privacy. Unauthorized access to this data could lead to breaches of confidentiality or misuse of personal information, underscoring the importance of implementing robust encryption protocols and access control mechanisms to mitigate these risks.

\noindent\textbf{Algorithmic efficiency and adaptability}: While the distributed Dijkstra's algorithm implemented on FPGA-based edge servers offers computational efficiency, ensuring accurate route calculations in dynamic urban environments remains a challenge. Optimizing algorithmic performance to account for changing traffic conditions, road closures, and unforeseen obstacles is essential to maintain response effectiveness.

\noindent\textbf{Human factors and training requirements}: Despite the automation of many aspects of emergency response, human factors and training requirements remain critical considerations. Ensuring that intervention teams are adequately trained to interpret and act upon guidance provided by the system, particularly in high-stress situations, is essential to maximize operational effectiveness and minimize response times.

\subsubsection{Perspective work}
\noindent\textbf{Prototype development and testing}: Initiating prototype development to validate the feasibility and functionality of the proposed architecture in controlled environments is a short-term objective. This will allow to test system components, evaluate performance metrics, and early identify potential hidden challenges.
    
\noindent\textbf{Integration of artificial intelligence}: Future research could explore the integration of artificial intelligence (AI) techniques, such as machine learning and predictive analytics, to enhance the intelligence and adaptability of the architecture. AI algorithms could analyze historical data patterns, anticipate potential emergencies, and dynamically adjust response strategies based on evolving conditions.

\noindent\textbf{Community engagement and feedback mechanisms}: Incorporating community engagement and feedback mechanisms into the architecture could foster greater collaboration between responders and the public. Crowd-sourced data and citizen reporting platforms could provide valuable insights into local conditions and facilitate more targeted response efforts.

\noindent\textbf{Interoperability and standardization}: Addressing interoperability challenges and promoting standardization across emergency response systems is critical to facilitate seamless communication and coordination between different stakeholders and technologies. Developing open standards and protocols could enhance interoperability and streamline integration efforts.

\noindent\textbf{Ethical and societal implications}: Further research is needed to explore the ethical and societal implications of deploying advanced emergency response technologies in urban environments. Understanding the potential impact on privacy, equity, and social cohesion is essential to ensure that the benefits of the architecture are equitably distributed and ethically implemented.

\noindent\textbf{Long-term evaluation and optimization}: Continuous evaluation and optimization of the architecture are necessary to address emerging challenges and evolving user needs. Long-term studies assessing the effectiveness, efficiency, and user satisfaction of the system in real-world settings can provide valuable insights for ongoing refinement and improvement.
%%%%%%%%%%%%%%%%%%%%%%%%%%%%%%%%%%%%%%%%%%%%%%%%%%%%%%%
%\input{discussion}
%%%%%%%%%%%%%%%%%%%%%%%%%%%%%%%%%%%%%%%%%%%%%%%%%%%%%%%
%%%%%%%%%%%%%%%%%%%%%%%%%%%%%%%%%%%%%%%%%%%%%%%%%%%%%%%
\section{Conclusion}
\label{sec:conclusion}
This paper presented a comprehensive architecture tailored to the unique challenges of emergency response systems in resource-constrained environments, exemplified by contexts akin to that of Chad. By leveraging decentralized computing resources and optimizing algorithms for efficiency, the proposed architecture addresses critical shortcomings observed in existing approaches. The integration of Field Programmable Gate Arrays (FPGAs), Dijkstra's algorithm, and Edge Computing enhances computational speed, scalability, and reliability, ensuring timely and effective emergency response.

The architecture incorporates key components such as the Surveillance and Alerting System (SAS), Edge Server Network (ESN), Intervention Order Processing System (IOPS), and Intervention Team Guidance System (ITGS). Together, these components form a holistic framework that significantly improves the timeliness and efficacy of emergency response operations. The SAS provides real-time situational awareness, enabling informed decision-making. The ESN leverages FPGA technology for fast computation of shortest paths, reducing response times. The IOPS optimizes resource allocation and deployment strategies, while the ITGS ensures continuous tracking and guidance during emergency interventions.

Our analysis and theoretical evaluations demonstrate the architecture's potential to drastically reduce response times and enhance resource allocation efficiency. By introducing parallel calculation of shortest paths, storing computed paths in cache memory and utilizing distributed edge computing, the system speeds calculations, minimizes redundant calculations and ensures reliable performance even in environments with constrained infrastructure and intermittent connectivity.

While this work represents a significant advancement in addressing the pressing needs of emergency response systems in resource-constrained environments, it is not without its limitations. The proposed architecture's performance relies heavily on the accuracy and timeliness of data collected from IoT devices and sensors. Future research endeavors should focus on refining algorithmic approaches, enhancing hardware capabilities, and conducting real-world implementations to validate the efficacy of the proposed architecture. Additionally, exploring machine learning techniques for predictive analytics and decision support could further enhance the system's capabilities.

In essence, this paper's contribution seeks to empower emergency responders with the tools and technologies necessary to mitigate the impact of crises and safeguard the well-being of communities in challenging environments. By fostering collaboration between academia, industry, and local stakeholders, we aspire to drive meaningful advancements in emergency response systems and contribute to the collective resilience of societies worldwide. The proposed architecture not only addresses current gaps in emergency response strategies but also sets a foundation for future innovations in smart city technologies, ultimately enhancing the safety and resilience of urban populations.

%%%%%%%%%%%%%%%%%%%%%%%%%%%%%%%%%%%%%%%%%%%%%%%%%%%%%%%

    % --- Final Declarations ---
    
    \begin{blindeddeclaration}{Authors' Information}
        \begin{itemize}[itemsep=-0.3em]
            \item \noindent\textbf{Mahamat Abdel Aziz Assoul}: Mahamat Abdel Aziz Assoul is a dedicated and aspiring scholar in the field of Industrial Engineering and Maintenance. He is making significant strides in his academic journey at the Polytechnic University of Mongo, located in Mongo, Chad. Currently pursuing his PhD in Industrial Engineering and Maintenance, Mahamat is under the expert guidance and supervision of Professor Abakar Mahamat Tahir and Professor Taibi Mahmoud.
    
            \item \noindent\textbf{Abakar Mahamat Tahir}: Professor Abakar Mahamat Tahir is a distinguished academic figure in the Department of Technical Sciences at the University of N'Djamena, situated in N'Djamena, Chad. With a profound commitment to advancing knowledge and expertise in the realm of technical sciences, Professor Tahir has made significant contributions to the field of thermal modeling in power electronics. In 2003, Professor Tahir obtained his PhD from the INSA school of Lyon, where he conducted groundbreaking research in the field of thermal modeling of magnetic components used in power electronics. Throughout his academic career, Professor Tahir has been committed to excellence in teaching, research, and scholarly inquiry. His expertise in thermal modeling has contributed to advancements in power electronics design, simulation, and optimization, with implications for various industrial applications.
            
            \item \noindent\textbf{Taibi Mahmoud}: Professor Taibi Mahmoud earned his PhD in Electronics on June 30, 2007. With a solid foundation in Electronics, he became a Professor in June 2014. His expertise led to him becoming a Research Director in October 2016. He currently serves as a Professor of Electronics at the Faculty of Engineering Sciences, Badji Mokhtar University, Annaba, Algeria. Professor Taibi's interests include Biological Engineering, Agro-food processes, and emerging estimation techniques such as Fuzzy Logic, Neural Networks, Genetic Algorithms, Hybrid Systems, and Hidden Markov Chains. He established his laboratory in 2008, focusing on Wireless Telecommunications, Embedded Communication Networks, Digital Communication, Information Security, Routing, and FPGA Implementation.
    
            \item \noindent\textbf{Garrik Brel Jagho Mdemaya}: Garrik Brel Jagho Mdemaya is presently pursuing a postdoctoral research fellowship at the University of Limpopo (South Africa) in the Department of Computer Science. Additionally, he holds the position of Assistant Lecturer in the Department of Computer Engineering at the Fotso Victor University Institute of Technology, University of Dschang (Cameroon). Dr. Jagho holds a Ph.D. obtained in 2022, a M.S. degree obtained in 2016, and a B.S. degree obtained in 2013, all specializing in computer science from the University of Dschang. His current research interests encompass parallel algorithms, distributed systems, wireless sensor networks, the Internet of Things, big data, cognitive radio networks, and edge computing.
    
            \item \noindent\textbf{Milliam Maxime Zekeng Ndadji}: Dr. Milliam Maxime Zekeng Ndadji is a researcher making contributions to the fields of computer science and software engineering. Holding a PhD in Computer Science obtained in 2021 from the University of Dschang, located in Dschang, Cameroon, Dr. Zekeng is currently serving as a lecturer and researcher in the Department of Mathematics and Computer Science at the University of Dschang. His research interests span a wide spectrum of topics within computer science. His expertise encompasses system architecture, software engineering methodologies, business process management, and AI-based software design and implementation. He is also making contributions to the advancement of formal methods models and languages, leveraging his expertise to develop novel approaches and techniques for software development and verification.
        \end{itemize}
    \end{blindeddeclaration}
        
    \begin{blindeddeclaration}{Authors' Contributions}
        \begin{itemize}[itemsep=-0.3em]
            \item \noindent\textbf{Mahamat Abdel Aziz Assoul}: Conceptualization, Methodology, Investigation, Writing - original draft, Writing - review and editing.

            \item \noindent\textbf{Abakar Mahamat Tahir}: Project administration, Conceptualization, Supervision, Validation.
            
            \item \noindent\textbf{Taibi Mahmoud}: Project administration, Conceptualization, Supervision, Validation.
    
            \item \noindent\textbf{Garrik Brel Jagho Mdemaya}: Identification of Research Problem, Conceptualization, Methodology, Investigation, Supervision, Validation, Writing - original draft, Writing - review and editing.
    
            \item \noindent\textbf{Milliam Maxime Zekeng Ndadji}: Identification of Research Problem, Conceptualization, Methodology, Investigation, Supervision, Validation, Writing - original draft, Writing - review and editing.
        \end{itemize}
    \end{blindeddeclaration}
        
    \begin{declaration}{Competing Interests}
        The authors declare that they have no competing interests.
    \end{declaration}
        
    \begin{declaration}{Funding}
        No funding was received for this project.
    \end{declaration}
        
    \begin{declaration}{Availability of Data and Material}
        The data used to support the findings of this study are available from the corresponding author upon request.
    \end{declaration}

    \references{bibliography}
\end{document}